\documentclass[conference]{IEEEtran}
\ifCLASSINFOpdf
\usepackage[pdftex]{graphicx}
\usepackage{array}
\graphicspath{{../pdf/}{../jpeg/}}
\DeclareGraphicsExtensions{.pdf,.jpeg,.png}
\else
\usepackage{algorithmic}
\usepackage{blkarray}
\usepackage{algorithm}
\usepackage[dvips]{graphicx}
\usepackage{notoccite}
\graphicspath{{../eps/}}
\DeclareGraphicsExtensions{.eps}
\fi
\begin{document}
\title{Parallel architectures for fuzzy triadic similarity learning}

\author{\IEEEauthorblockN{Sonia Alouane Ksouri$^1$, Minyar Sassi Hidri$^2$, Kamel Barkaoui$^3$}
\IEEEauthorblockA{$^{1,2}$Universit\'e Tunis El Manar\\ $^{1,2}$Ecole Nationale d'Ing\'enieurs de Tunis\\Laboratoire Signal, Image et Technologies de l'Information\\
BP. 37, Le Belv\`ed\`ere 1002, Tunis, Tunisia\\
$^{3}$CEDRIC-CNAM\\Rue Saint-Martin Paris 75003, France\\
$\{^{1,2}$sonia.alounae,minyar.sassi$\}$@enit.rnu.tn,$^3$kamel.barkaoui@cnam.fr}
}
\maketitle

\begin{abstract}
%\boldmath
In a context of document co-clustering, we define a new similarity measure which iteratively computes similarity while combining fuzzy sets in a three-partite graph. The fuzzy triadic similarity (FT-Sim) model can deal with uncertainty offers by the fuzzy sets. Moreover, with the development of the Web and the high availability of storage spaces, more and more documents become accessible. Documents can be provided from multiple sites and make similarity computation an expensive processing. This problem motivated us to use parallel computing. In this paper, we introduce parallel architectures which are able to treat large and multi-source data sets by a sequential, a merging or a splitting-based process. Then, we proceed to a local and a central (or global) computing using the basic FT-Sim measure. The idea behind these architectures is to reduce both time and space complexities thanks to parallel computation.

\end{abstract}


% \fi
\IEEEpeerreviewmaketitle

\section{Introduction}
Nowadays information on the internet is exploding exponentially through time, and approximately 80\%
are stored in the form of text. So text mining has been a very hot topic. One particular research area
is document clustering, which is a major topic in the Information Retrieval community. It allows to efficiently capture high-order similarities between objects described by rows and columns of a data matrix. In the domain of text clustering, a document is described as a set of words.

The relationship between documents and words allows for exploitation of the relationship between groups of words that occur mostly in a group of documents.

In \cite{X-Sim}, a co-similarity measure has been proposed, called X-Sim \cite{X-Sim} which builds on the idea of iteratively generating the similarity matrices between documents and words, each of them built on the basis of the other. This measure works well for unsupervised document clustering.

However, in recent researches, the sentence has been considered as a more informative feature term for improving the effectiveness of document clustering \cite{phrase}. While considering three levels Documents $\times$ Sentences $\times$ Words to represent the data set, we are able to deal with a dependency between Documents-Sentences, as also between Sentences-Words and, by deduction, between Documents-Words.

Another important aspect in co-clustering is the weight computing. A weighted value may be assigned as a link from a document to a word (or sentence) indicating the presence of the word (sentence) in that document. The 0/1 encoding denotes the presence or absence of an object in a given document.

Different weighting schemes such as the tf-idf \cite{tf-idf} may be incorporated to better represent the importance of words in the corpus, but it has spawned the view that classical probability theory is unable to deal with uncertainties in natural language and machine learning.

So, we proceed to a fuzzification control process which converts crisp similarities to fuzzy ones. The conversion to fuzzy values is represented by the membership functions \cite{kundi}. They allow a graphical representation of a fuzzy set \cite{zadeh1965}. These fuzzy similarity matrices are used to calculate fuzzy similarity between documents, sentences and words in a triadic computing called FT-Sim (Fuzzy Triadic Similarity).

Moreover, with the development of the Web and the high availability of the storage spaces, more and more documents become accessible. Data can be provided from multiple sites and can be seen as a collection of matrices. By separately processing these matrices, we get a huge loss of information.

Several extensions to the co-clustering methods have been proposed to deal with such multi-view data. Some works aim at combining multiple similarity matrices to perform a given learning task \cite{multi-graph,multi-matrices}. The idea being to build clusters from multiple similarity matrices computed along different views.

Multi-view co-clustering such as MV-Sim \cite{MV-Sim} architecture, based on X-Sim measure \cite{X-Sim} deals with the problem of learning co-similarities from a collection of matrices describing interrelated types of objects. It was proved that this architecture provides some interesting properties both in terms of convergence and scalability and it allows an efficient parallelization of the process.

For this, we provide parallel architectures for FT-Sim  to tackle the problem of learning similarities from a collection of matrices. For multi-source or large matrices, we propose different parallel architectures in which each FT-Sim is the basic component or node we will use to deal with multiple matrices.

Thus, we consider a model in which data sets are distributed into $N$ sites (or relation matrices). They describe the connections between documents for each local data set.

Our goal is then to compute a fuzzy Documents $\times$ Documents matrix $\widetilde{D}^{(i)}_{2}$ for each site $i$ $(i=1..N)$ trying to take into account all the representative information expressed in the relations.

To combine multiple occurrences of FT-Sim, we propose sequential, merging and splitting based parallel architectures.

The rest of the paper is organized as follows: in section 2 we highlights backgrounds related to similarity measures in a multi-view data sets. In section 3 we provide our fuzzy triadic similarity measure. In section 4 we present the three proposed architectures allowing parallel computing for co-clustering. Section 5 concludes the paper and gives indications of some future work.
\section{Dealing with Multi-view Data sets}
Most of the existing clustering methods focus on data sets described by a unique data matrix, which can either be a matrix which describes objects by their characteristics, or a relation matrix that describes the intensity of the relation between instances of two types of objects, such as a Documents $\times$ Words matrix. In the latter case, both types of objects can be clustered ; methods dealing with this task are referred to as co-clustering approaches and have been extensively studied.

However, in many applications, data sets involving more than two types of interacting objects, or simply related, are also frequent. A simple way to represent such data sets is to use as many matrices as there are relations between the objects. Then, one could use classical co-clustering methods to separately cluster the objects occurring in the different matrices but, in this way, interactions between objects are not taken into account, thus leading to a loss of information. Therefore, handling the views together, referenced as the multi-view clustering task, is an interesting challenge in the learning domain to resolve limits of classical clustering.

Many extensions to the clustering methods have been proposed to deal with multi-view data. In \cite{multi-view}, they  describe an extension of k-means (MVKM) and of EM algorithms using multi-view model. In \cite{multi-graph} and \cite{multi-matrices}, the authors build clusters from multiple similarity matrices computed along different views. In \cite{co-training}, a co-clustering system called MVSC has been proposed. It permits a multi-view spectral clustering while using the co-training that has been widely used in semi-supervised learning problems. The general idea is to learn the clustering in one view and use it to label the data in an other view so as to modify the graph structure (similarity matrix).

Closer to our approach, some works aim at combining multiple similarity matrices to perform a given learning task. The MVSim architecture \cite{MV-Sim} which is an extension of the X-Sim algorithm \cite{X-Sim}, adapts the previous algorithm to the multi-view context. It computes simultaneously the co-similarity matrix for each of $N$ different kinds of objects $T_i$ described by $M$ relation matrices. The basic idea is to create a learning network isomorphic to these data sets structures. It was shown that it is possible to use this architecture to efficiently compute co-similarities on large data sets by splitting a data matrix into smaller ones.
\section{FT-Sim: Fuzzy Triadic Similarity}
Sentence-based analysis means that the similarity between documents should be based on matching sentences rather than on matching single words only. Sentences contain more information than single words (information regarding proximity and order of words) and have a higher descriptive power\cite{phrase-simi} \cite{Cortex}\cite{bigr-trigram}. Thus a document must be broken into a set of sentences, and a sentence is broken into a set of words. We focus on how to combine the advantages of two representation models in document co-clustering.

To represent our textual data set, two representations have been proposed: the collection of matrices and the k-partite graph \cite{kpartite}. In the first, each matrix describes a view on the data. In the second, a graph is said to be k-partite when the nodes are partitioned into $k$ subsets with the condition than no two nodes of the same subset are adjacent. Thus in the k-partite graph paradigm \cite{kpartite}, a given subset of nodes contains the instances of one type of objects, and a link between two nodes of different subsets represents the relation between these two nodes.

To explain our model we consider matrices to represent the data sets and we use a three-partite graph representation of the data matrices with three relations linking to explain our model.

\begin{figure}[h!]
\centering
\includegraphics[height=4.5cm]{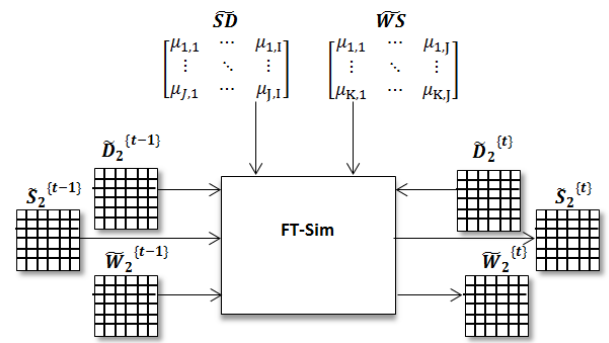}
\caption{Functional diagram of FT-Sim.}
\label{fig:1}
\end{figure}

From a functional point of view, the proposed FT-Sim model can be represented in the following way as shown in figure \ref{fig:1}, where $SD$ and $WS$ are two data matrices representing a corpus and describing the connection between Documents/Sentences and Sentences/Words, brought by the three-partite graph \cite{icwit}.

After the generation of $SD$ and $WS$ matrices, we proceed to a fuzzification process. It converts crisp values to fuzzy ones. The conversion to fuzzy values is represented by the membership functions \cite{kundi}. They allow a graphical representation of a fuzzy set \cite{zadeh1965}. There are various methods to assign membership values or the membership functions to fuzzy variables. We mention essentially the triangular and trapezoidal ones. The second form is the most suitable one for modeling fuzzy Sentences $\times$ Documents and Words $\times$ Sentences similarities.

For each document, we define a fuzzy membership function through a linear transformation between the lower bound value $L_{i}$, a membership of $0$, to the upper bound value $U_{i}$, which is assigned a membership of $1$. This function is used because smaller values linearly increase in membership to the larger values for a positive slope and opposite for a negative slope.

The following formulas show the fuzzy linear membership functions for $\widetilde{SD}_{i}$ and $\widetilde{WS}_{j}$.
\begin{equation}
\scriptsize{
\widetilde{SD}_i= [\mu]_{ji} = \left\{
\begin{array}{l l}
1 , & \qquad \qquad if~~SD_{ji}\geq L_{i}\\

\frac{SD_{ji}-U_{i}}{U_{i}-L_{i}} , & \qquad \qquad if~~L_{i} < SD_{ji}<U_{i} \\
0 , & \qquad \qquad if~~SD_{ji} \leq L_{i}\\
 \end{array} \right.
}
\end{equation}

and
\begin{equation}
\scriptsize{
\widetilde{WS}_j= [\mu]_{kj} = \left\{
\begin{array}{l l}
1 , & \qquad \qquad if~~WS_{kj}\geq L_{i}\\

\frac{WS_{kj}-U_{i}}{U_{i}-L_{i}} , & \qquad \qquad if ~~L_{i} < WS_{kj} <U_{i} \\
0 , & \qquad \qquad if~~WS_{kj} \leq L_{i}\\
 \end{array} \right.
 }
 \end{equation}

Before proceeding to fuzzy triadic computing, we must initialize Documents $\times$ Documents, Sentences $\times$ Sentences and Words $\times$ Words fuzzy matrices with the identity ones denoted as $\tilde{D}_2^{(0)}$, $\tilde{S}_2^{(0)}$ and $\tilde{W}_2^{(0)}$. The similarity between the same documents (resp. sentences and words) have the value equal to 1. All others values are initialized with zero. $\tilde{D}_2^{(t)}$ is as follows:

\begin{equation}
\scriptsize{
\tilde{D}_2^{(t)}= [\mu]_{lm}^{(t)}=~\begin{blockarray}{cccc}
  D_1  & \dots & D_m  \\
  \begin{block}{[ccc]c}
  1 & \dots  & \mu_{1m}^{(t)} & ~D_{1}  \\
  \vdots & \ddots & \vdots &      \\
  \mu_{l1}^{(t)} & \dots  & 1 & ~D_{l}  \\
  \end{block}
\end{blockarray}
}
\end{equation}

where $\mu_{lm}^{(t)}$ $(l=1..I$, $m=1..I$ is the membership degree of the $l^{th}$ document according the $m^{th}$ one. Similarly, we determine the $\tilde{S}_2^{(t)}$ and $\tilde{W}_2^{(t)}$.

After initializing $\tilde{D}_2^{(t)}$, we calculate the new matrix $\tilde{D}_2^{(t)}$ which represents fuzzy similarities between documents while using $\tilde{S}_2^{(t-1)}$ and $\widetilde{SD}$.

Usually, the similarity measure between two documents $D_l$ and $D_m$ is defined as a function that is the sum of the similarities between shared sentences.

Our idea is to generalize this function in order to take into account the intersection between all the possible pairs of sentences occurring in documents $D_l$ and $D_m$. In this way, not only can we capture the fuzzy similarity of their common sentences but also the fuzzy ones coming from sentences that are not directly common in the documents but are shared with some other documents.For each pair of sentences not directly shared by the documents, we need to take into account the fuzzy similarity between them as provided by $\tilde{S}_2^{(t-1)}$.

Since we work with fuzzy matrices formed by membership degrees, we should certainly be applied in accordance with the operators for fuzzy sets, especially the intersection and union. Thus, $\mu_{l,m}^{(t)}$, except the case $l=m$, can be formulated as follows:
\begin{equation}
\scriptsize{
\mu_{l,m}^{(t)}=\sum_{i=1}^{J}\sum_{j=1}^{J}\min(\mu_{il},\mu_{jm})\ast \mu_{ij}^{\tilde{S_2}^{(t-1)}}
}
\end{equation}

As we have shown for $\tilde{D}_2^{(t)}$ computing, we generalize fuzzy similarities in order to take into account the intersection between all the possible pairs of words occurring in sentences $S_l$ and $S_m$. In this way, not only do we capture the fuzzy similarity of their common words but also the fuzzy ones coming from words that are not directly common in the sentences but are shared with some other sentences.

For each pair of words not directly shared by the sentences, we need to take into account the fuzzy similarity between them as provided by $\tilde{W}_2^{(t-1)}$. The overall fuzzy similarity between documents $S_l$ and $S_m$ is defined in the following equation:
\begin{equation}
\scriptsize{
\mu_{l,m}^{(t)}=Min[\sum_{i=1}^{I}\sum_{j=1}^{I}\min(\mu_{il},\mu_{jm})\ast \mu_{ij}^{\tilde{D_2}^{(t-1)}},
}
\end{equation}
\begin{center}
\scriptsize{
$~\sum_{i=1}^{K}\sum_{j=1}^{K}\min(\mu_{il},\mu_{jm})\ast \mu_{ij}^{\tilde{W_2}^{(t-1)}}]$
}
\end{center}

Similarly, for each pair of words not directly shared by the sentences, we need to take into account the fuzzy similarity between them as provided by $\widetilde{W}_2^{(t-1)}$. The overall fuzzy similarity between documents $W_l$ and $W_m$ is defined in the following equation:
\begin{equation}
\scriptsize{
\mu_{l,m}^{(t)}=\sum_{i=1}^{J}\sum_{j=1}^{J}\min(\mu_{il},\mu_{jm})\ast \mu_{ij}^{\tilde{S_2}^{(t-1)}}
}
\end{equation}
\section{Parallel FT-Sim}
For multi-source or large data sets, we propose different parallel architectures in which each FT-Sim is the basic component or site we will use to deal with multiple matrices.

Thus, we consider a model in which the data sets are composed of $N$ relation matrices $\widetilde{SD}^{(i)}$ and $\widetilde{WS}^{(i)}$ $(i=1..N)$. They describe the connections between documents for each local data set. Our goal is then to compute a fuzzy matrix $\widetilde{D}^{(i)}_{2}$ for each data set trying to take into account all the information expressed in the relations.

To combine multiple occurrences of FT-Sim, we can adopt three different architectures: a sequential, a merging or a splitting based one.
\subsection{Sequential-based parallel architecture}
\begin{figure*}
\centering
\includegraphics[height=4.7cm]{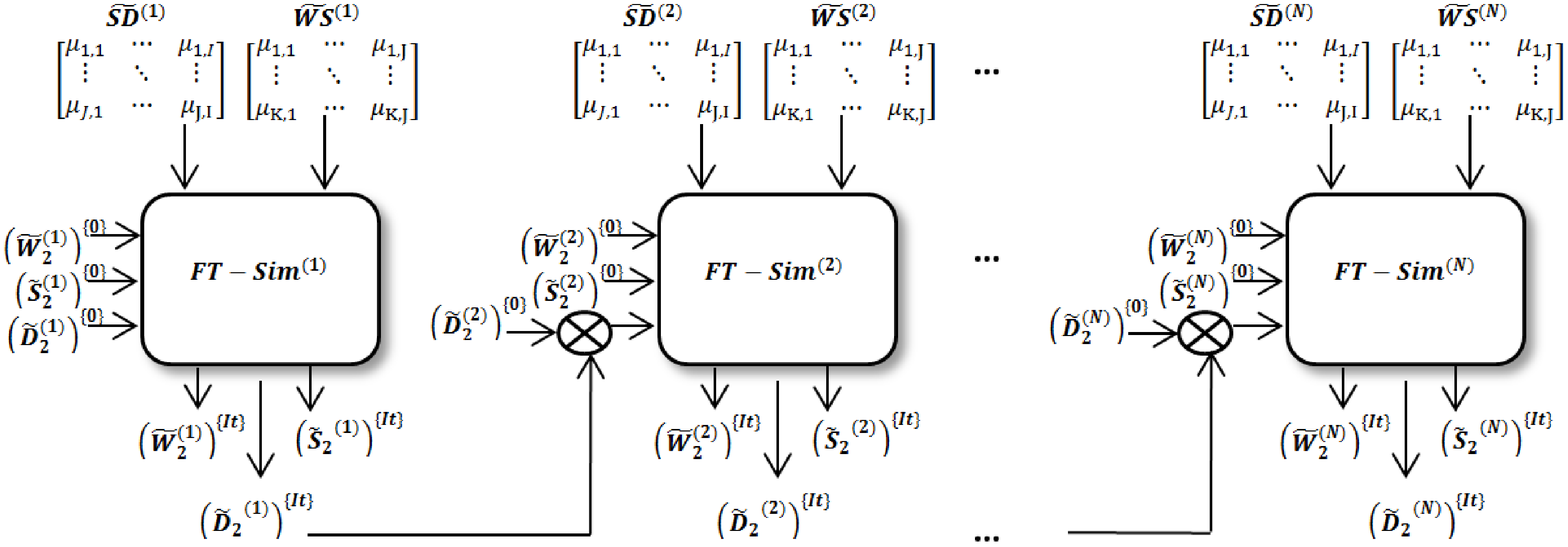}
\caption{Sequential-based parallel architecture.}
\label{fig:2}
\end{figure*}

In this first model, an instance of $FT-Sim^{(i)}$ is associated to each local site $i$. Each site is represented by the relation matrice corresponding to the similarity between sentences/documents $\widetilde{SD}^{(i)}$ and  words/sentences $\widetilde{WS}^{(i)}$ for $(i = 1..N)$. $N$ being the number of data sources. This instance is denoted $FT-Sim^{(i)}$. Figure \ref{fig:2} shows the sequential-based parallel architecture.

As shown in figure \ref{fig:2}, we assume a link between each $FT-Sim^{(i)}$ and the following one. Then it computes the similarity matrices from the data matrices of the first data set $\widetilde{SD}^{(1)}$ and  $\widetilde{WS}^{(1)}$, and uses the resulting document similarity matrix to initialize the next site.

The document similarity issue of the $1^{(st)}$ data-set $\widetilde{D}_2^{(1)}$ is used to initialize the next document similarity denoted by $(\widetilde{D}_2^{(2)})^{\{0\}}$ (the second document similarity matrice at iteration $0$). The initialization function presented in algorithm 1 is then run with a second $\widetilde{SD}^{(2)}$ and $\widetilde{WS}^{(2)}$ matrices etc.

\begin{algorithm}%[ht!]
\footnotesize{
\caption{Initialization function}
\label{alg1}
\begin{algorithmic}[1]
\REQUIRE $\widetilde{D}_2^{(i)}$
\ENSURE $\tilde{D}_2$
\STATE $I\leftarrow$ Compute the number of documents in $\widetilde{D}_2^{(i)}$ and $(\widetilde{D}_2^{(i+1)})^{\{0\}}$
\STATE Let $\tilde{D}_2 =[\mu_{l,m}]~~(l=1..I$ and $m=1..I)$
\STATE $\tilde{D}_2\leftarrow$ $Identity$
\FOR {$l=1..I^{(i)}$}
\FOR {$m=1..I^{(i)}$}
\STATE $\mu_{l,m}\leftarrow \mu_{l,m}^{(i)}$
\ENDFOR
\ENDFOR
\end{algorithmic}
}
\end{algorithm}

The natural question that arises is: how to initialise $(\widetilde{D}_2^{(i+1)})^{\{0\}}$ with $\widetilde{D}_2^{(i)} $?

In the beginning, $(\widetilde{D}_2^{(i+1)})^{\{0\}}$ must contain all documents existing in the $i^{th}$ and the $(i+1)^{th}$ data sets. They are initialized as an identity matrix denoted by $Identity$.

After that, the obtained $(\widetilde{D}_2^{(i+1)})^{\{0\}}$ is updated with the similarities in  $\widetilde{D}_2^{(i)}$. The different steps for the sequential-based parallel process are presented in algorithme 2.

\begin{algorithm}%[ht!]
\footnotesize{
\caption{Sequential-based algorithm}
\label{alg2}
\begin{algorithmic}[1]
\REQUIRE $\widetilde{D}_2^{(0)}$, $N$
\ENSURE $\tilde{D}_2$
\STATE Execute $FT-Sim^{(1)}$ with $\tilde{D}_2^{(0)}$
\FOR {$i=1..N$}
\STATE Initializing with $\widetilde{D}_2^{(i)}$
\STATE Execute $FT-Sim^{(i+1)}$ with $\widetilde{D}_2^{(i)}$
\ENDFOR
\end{algorithmic}
}
\end{algorithm}

Each $FT-Sim^{(i)}$ is connected to the inputs of the following one which creates a chain. In that way, the instances are sequentially run in a static or dynamic order and the similarity matrices $\widetilde{D}_2^{(i)}$ are progressively updated.

The problem with this model is that the order matters. How do we choose the order of the matrices? How many iterations do we perform for each local $FT-Sim^{(i)}$?

Thus, without any prior knowledge about the relative interest of the relation matrices and the number of iterations for each local computing, this model seems difficult to optimize.
\subsection{Merging-based parallel architecture}
\begin{figure*}
\centering
\includegraphics[height=7.6cm]{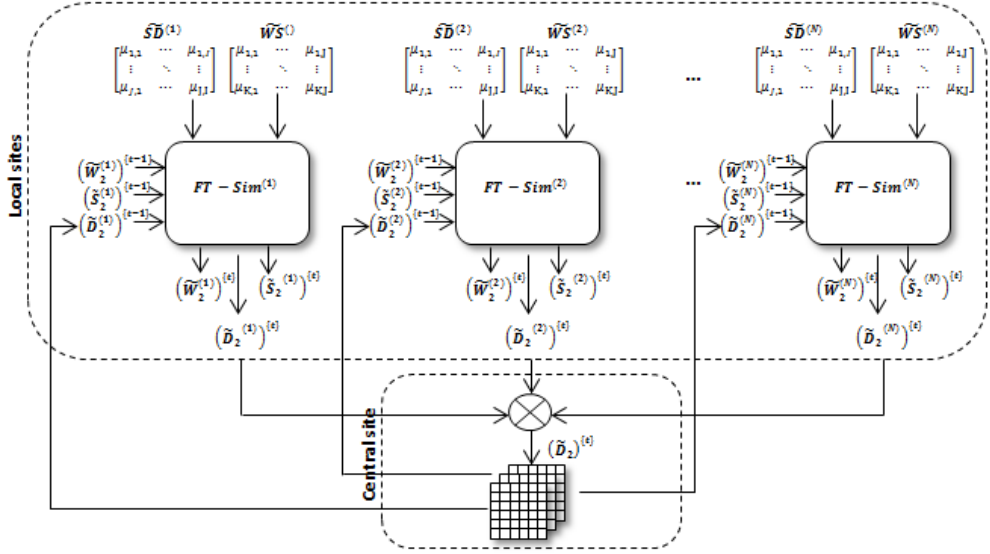}
\caption{Merging-based parallel architecture.}
\label{fig:3}
\end{figure*}

In the second model, we propose to compute the similarity matrices from several sites and merge them before performing the co-clustering algorithm on it. Figure \ref{fig:3} shows the merging-based parallel architecture.

In this topology, all local $FT-Sim^{(i)}$ instances $(i = 1..N)$ are run in parallel, then the similarity matrices $\widetilde{D}_2^{(i)}$ are simultaneously updated with an aggregation function. This policy offers the benefit that all the instances of $FT-Sim^{(i)}$ have the same influence.

The aggregation function takes $N$ matrices $(\widetilde{D}_2^{(1)})^{\{t\}}$, $(\widetilde{D}_2^{(2)})^{\{t\}}$,..,$(\widetilde{D}_2^{(N)})^{\{t\}}$ issue from each data source $i$ for a given iteration $t$. Two rules are adopted:

\textit{Rule 1:} If a given document does not appear in a single site then we assign its corresponding similarity measures directly in $\tilde{D}_2$.

\textit{Rule 2:} If a particular document appears in several different sites, we assign the minimum of all similarity measures relevant to this document to $\tilde{D}_2$ without taking into account the value of 0.

The different steps of aggregation computing are presented in algorithm 3.

\begin{algorithm}%[ht!]
\footnotesize{
\caption{Merging Function}
\label{alg3}
\begin{algorithmic}[1]
\REQUIRE Collection of $N$ matrices $\{(\widetilde{D}_2^{(1)})^{\{t\}}$, $(\widetilde{D}_2^{(2)})^{\{t\}}$, .., $(\widetilde{D}_2^{(N)})^{\{t\}}\}$
\ENSURE $\widetilde{D}_2$
\STATE $I\leftarrow$ Number of documents in $\{(\widetilde{D}_2^{(1)})^{\{t\}}$, $(\widetilde{D}_2^{(2)})^{\{t\}}$, .., $(\widetilde{D}_2^{(N)})^{\{t\}}\}$
\STATE Let $\widetilde{D}_2 = [\mu_{l,m}]$ $(l=1..I$ and $m=1..I)$
\STATE $\widetilde{D}_2\leftarrow Identity$
\FOR {Each document $D_l$ of $\widetilde{D}_2$ }
\IF{$D_l$ Appear in only one data set $s$}
\STATE $\mu_{l,*}\leftarrow \mu_{l,*}^{(s)}$
\ELSE
\STATE $\mu_{l,*}\leftarrow \min($ All$ \mu_{l,*}^{(i)}$) $i\in \{$ sites where $D_l$ appear$\}$ with $\mu_{l,*}^{(i)}\neq 0$
\ENDIF
\ENDFOR
\end{algorithmic}
}
\end{algorithm}

So, for a given iteration $t$, each instance $FT-Sim^{(i)}$ produces its own similarity matrix $(\widetilde{D}_2^{(i)})^{\{t\}}$. We thus get a set of output similarity matrices $\{(\widetilde{D}_2^{(1)})^{\{t\}}$, $(\widetilde{D}_2^{(2)} )^{\{t\}}$,..,$(\widetilde{D}_2^{(N)} )^{\{t\}}\}$ the cardinal of which being equal to the number of data-sets related to $N$.

Therefore, we use the aggregation function denoted by $\bigotimes$ and developed in the merging based function to compute a consensus similarity matrix merging all of the $\{(\widetilde{D}_2^{(1)})^{\{t\}}$, $(\widetilde{D}_2^{(2)})^{\{t\}}$,..,$(\widetilde{D}_2^{(N)})^{\{t\}}\}$ with the current matrix $\widetilde{D}_2^{\{t\}}$.

In turn, this resulting consensus matrix is connected to the inputs of all the $FT-Sim^{(i)}$ instances, to be taken into account in the ${t+1}^{th}$ iteration, thus creating feedback loops allowing the system to spread the knowledge provided by each $(\widetilde{D}_2^{(i)})^{\{t\}}$ within the network. The different steps for the merging-based
parallel process are presented in algorithme 4.

\begin{algorithm}%[ht!]
\footnotesize{
\caption{Parallel merging-based Algorithm}
\label{alg4}
\begin{algorithmic}[1]
\REQUIRE collection of matrices $\widetilde{SD}^{(i)}$, $\widetilde{WS}^{(i)}$ $(i=1..N)$, $T$
\ENSURE $\widetilde{D}_2$
\FORALL {$i$}
\STATE $(\widetilde{D}_2^{(i)})^{\{0\}}\leftarrow Identity$
\STATE $(\widetilde{S}_2^{(i)})^{\{0\}}\leftarrow Identity$
\STATE $(\widetilde{W}_2^{(i)})^{\{0\}}\leftarrow Identity$
\FOR {$i=1..T$}
\STATE Execute every $FT-Sim^{(i)}$ with $\widetilde{SD}^{(i)}$, $\widetilde{WS}^{(i)}$ and $t=1$
\STATE $(\widetilde{D}_2)^{\{t\}}\leftarrow$ Merging with all $(\widetilde{D}_2^{[i)})^{\{t\}}$
\STATE Update each $(\widetilde{D}_2^{[i)})^{\{t\}}$
\ENDFOR
\ENDFOR
\end{algorithmic}
}
\end{algorithm}

The complexity of this architecture is obviously related to that of the $FT-Sim^{(i)}$ algorithm. In the parallel merging-based architecture, as each instance of $FT-Sim^{(i)}$ can run on an independent core, the method can easily be parallelized, thus keeping the global complexity unchanged (considering the number of iterations as a constant factor). So, the complexity of the merging function can be ignored.
\subsection{Splitting-based parallel architecture}
\begin{figure*}
\centering
\includegraphics[height=6.2cm]{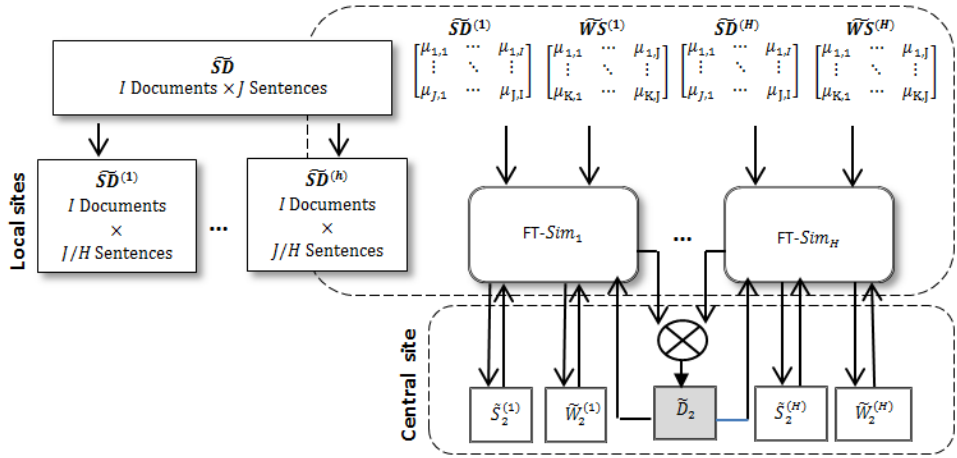}
\caption{Splitting-based parallel architecture.}
\label{fig:4}
\end{figure*}

In this section we present a generated model that can use previous architectures to efficiently compute FT-Sim on large data sets by splitting a data matrix into smaller ones. Figure \ref{fig:4} shows the splitting-based parallel architecture.

In order to reduce the complexity of a problem of treating huge data sets, it is possible to split a given data matrix  into a collection of smaller ones, each sub-matrix becoming a component of our network and processed as a separate view.

We have to evaluate the splitting approaches with the aim of finding the one most suitable with our solution. Here, our goal is to cluster the documents and to explore the behavior of the proposed architecture when varying the number of $H$ splits, obtaining $H$ sub-matrices. Then we adopt a random split sentence method. For each $SD^{(i)}$ matrix, the sentences are divided into $H$ sub-sets thereby forming $H$ sub-matrices $SD^{(i)}$. So, The number of $FT-Sim^{(i)}$ instances in the proposed network is equal to the number of splits $H$.

For example, let us consider a problem with one [documents/sentences] matrix of size $I$ by $J$ in which we just want to cluster the documents. we can divide the problem into a collection of $h$ matrices of size $n$ by $m/H$. Thus, by using a distributed version of $FT-Sim^{(i)}$ on $h$ cores, we will gain both in time and space complexity.

By splitting a matrix, we lost some information. The solution does not compute the co-similarities between all pairs of sentences but only between the words occurring in each $\widetilde{SD}^{(i)}$. Thanks to the feedback loops of this architecture and to the presence of the common similarity matrix $\tilde{D}_2$, we will be able to spread the information through the network and alleviate the problem of inter-matrice comparisons.

Thus, by using a parallel version of $FT-Sim^{(i)}$ on $H$ cores, we will gain both in time and space complexity: indeed,the time complexity decreases, leading to an overall gain of $1/H^2$ \cite{MV-Sim-complexite}. In the same way, the memory needed to store the similarity matrices between words will decrease by a $1/H$ factor.
\section{Conclusion}
In this paper, a fuzzy triadic similarity model, called FT-Sim, for the co-clustering task has been proposed. It takes, iteratively, into account three abstraction computing levels Document $\times$ Sentences $\times$ Words. The sentences consisting of one or more words are used to designate the fuzzy similarity of two documents. We are able to cluster together documents that have similar concepts based on their shared (or similar) sentences and in the same way to cluster together sentences based on words. This also allows us to use any classical clustering algorithm such as Fuzzy-C-Means (FCM) \cite{bezdek1984} or other fuzzy partitioned-based clustering approaches \cite{MacQueen1967}.

Our proposition has been extended to suit with multi-view models. Because the domain of text clustering focuses on documents and their similarities, in our proposition  we spread informations about document similarities. We have presented three parallel architectures that combine FT-Sim instances to compute similarities from different sources.

Actually, we need to further analyze the theoretical points of view and the behavior of the three architectures in a multi-threading programming.

\end{document}